\documentclass[aps, pre, groupedaddress, twocolumn, floatfix, showkeys, notitlepage, nofootinbib, 10pt]{revtex4-1}
\usepackage{amsmath,amssymb}
\usepackage[utf8]{inputenc}
\usepackage[pdftex]{graphicx}	

\newcommand{\ignore}[1]{}

\newcommand{\half}{\frac{1}{2}}
\newcommand{\dhalf}{\dfrac{1}{2}}

\newcommand{\bigOh}[1]{\mathcal{O}\left(#1\right)}
\newcommand{\Bwedge}{\mathbf{B}_\wedge}
\newcommand{\Btriangle}{\mathbf{B}_{\triangle}}

\begin{document}
\title{Transitions from trees to cycles in adaptive flow networks} 
\author{Erik Andreas Martens\,$^{1,2,3,*}$}
\email{eama@dtu.dk}
\author{Konstantin Klemm\,$^{4,*}$}
\email{klemm@ifisc.uib-csic.es}
\affiliation{
$^{1}$ Department of Applied Mathematics and Computer Science, Technical University of Denmark, 2800 Kgs. Lyngby, Denmark\\
$^{2}$ Department of Biomedical Sciences, University of Copenhagen, Blegdamsvej 3, 2200 Copenhagen, Denmark\\
$^{3}$ Department of Mathematical Sciences, University of Copenhagen, Universitetsparken 5, 2200 Copenhagen, Denmark\\
$^{4}$ IFISC (CSIC-UIB), Campus Universitat de les Illes Balears, Palma de Mallorca, Spain
}

\begin{abstract}
Transport networks are crucial to the functioning of natural and
technological systems. Nature features transport networks that are
adaptive over a vast range of parameters, thus providing an impressive
level of robustness in supply. Theoretical and experimental studies have
found that real-world transport networks exhibit both tree-like motifs
and cycles. When the network is subject to load fluctuations, the
presence of cyclic motifs may help to reduce flow fluctuations and,
thus, render supply in the network more robust. While
previous studies considered network topology via optimization
principles, here, we take a dynamical systems approach and study a
simple model of a flow network with dynamically adapting weights
(conductances). We assume a spatially non-uniform distribution of rapidly fluctuating loads in the
sinks and investigate what network configurations are dynamically
stable. The network converges to a spatially non-uniform stable
configuration composed of both cyclic and tree-like structures. Cyclic
structures emerge locally in a transcritical bifurcation as the
amplitude of the load fluctuations is increased. The resulting adaptive
dynamics thus partitions the network into two distinct regions with
cyclic and tree-like structures. The location of the boundary between
these  two regions is determined by the amplitude of the fluctuations.
These findings may explain why natural transport networks display cyclic
structures in the micro-vascular regions near terminal nodes, but
tree-like features in the regions with larger veins.
\end{abstract}

\maketitle

\section{Introduction}

Network structures are found in all of our everyday life, ranging from social interactions over technological infrastructure to natural systems. Networks serve vital functions on microscopic to macroscopic length scales, ranging from proteins, DNA, cells, organs and organisms~\cite{Strogatz2001}. An important function of networks is to transport people, goods, metabolites, and information among other~\cite{Hufnagel2004a,Brockmann2006,BarabasiOltvai2004,Kaluza2010,travers1967small,Marbach2016}. While in technology, transport networks are relatively rigid and yield limited adaptivity,  biological transport networks are capable of ensuring robust flow and operate satisfactorily over a vast range of parameters to prevent operational failure even under extreme conditions.

An example for one of the most advanced transport networks is the mammalian vascular network. Every second, the vasculature is without interruptions serving regions of the brain despite ever changing neural activity~\cite{Devor2007} or changing demands in other tissue~\cite{Jacobsen2009}.  
The mammalian vasculature, composed of bifurcations (nodes) and vessels (links), is highly adaptive because vessel diameters (weights) dynamically adjust to changes in flow properties such as pressure and shear stress via a variety of vessel response mechanisms~\cite{Jacobsen2009}. Several biophysical models have investigated the response mechanism of \emph{single} vessels~\cite{Feldberg1995,Alstrom1999,Jacobsen2003,vanBavelTuna2014}. Here, we aim at understanding how an adaptive flow network may respond on a \emph{global network level}. Notably, the mammalian vasculature forms a complex network~\cite{Kassab1993,Reichholdt2009} displaying both tree-like~\cite{Kassab2006} and cyclic motifs~\cite{Blinder2013} which also are found in the leaves of trees.

Recent work investigated optimal topology of flow networks from the perspective of their energy efficiency, damage resilience, or cost of repair~\cite{Dodds2010,Corson2010,Rubido2014,Farr2014,Tero2010}, features which may be argued to have evolved over long time scales. From the perspective of optimization theory, it is interesting to note that depending on the shape of the cost function, tree-like or cyclic structures may be more effective~\cite{Kantorovich1942translocation,Villani2003,Bohn2007,Durand2007}, leading to phase transitions between tree-like and cyclic structures. Moreover, cycles not only confer redundant structures and thus improve damage resilience (i.e., another cost function), but may also be more favorable in the presence of fluctuations~\cite{Corson2010,Katifori2010}.

However, while research on adaptive (co-evolving) networks from the perspective of theoretical physics is under active development, the understanding of adaptive \emph{flow} networks in particular remains relatively un(der)explored in the field of network theory~\cite{Gross2008}.
Here, we consider the dynamic stability of particular network configurations given that vessels may slowly adapt their diameters slowly over relatively short time scales. 
Specifically, we wish to address the following questions: given a flow network with dynamically adapting weights, (i) what network configurations (weights) are dynamically stable, and (ii) if non-uniform flow (load) fluctuations are present in the network, how far do these fluctuations affect adaptation into network regions where fluctuations are absent? In other words, when spatially inhomogeneous load fluctuations are present, does the network partition into clusters exhibiting tree-like motifs and cycles?

We wish to establish a fundamental understanding of the possible dynamics and gain insights from the perspective of network theory, nonlinear dynamics and physics, rather than of physiological specific aspects. Thus, to obtain answers to the above questions, we dispose of the mathematical intricacies inherent to solving biophysically detailed models. Instead of building on physiological models of blood vessel changes which occur via acute responses tone or the slower remodeling of the vessel~\cite{Hacking1996,Feldberg1995,Alstrom1999,Jacobsen2003,vanBavelTuna2014,Lee2006,Postnov2016}, we defer to simple conceptual models of adaptive flow networks, based on basic physical principles, allowing for analytical tractability.

We assume that the network has one constant inlet (source) and many outlets (sinks) subject to load fluctuations. In the language of vascular physiology, we model a bifurcating arterial network where the only inlet is a feeding artery, bifurcating in a tree-structure to the terminal nodes interfacing via capillaries to the veinous network. At this interface, changing supply demands constitute load fluctuations which are rapid compared to the adaptive network dynamics. 
A similar (though physiologically different) situation is seen in (real) trees, where the stem feeds the tree  with water and nutrients at a more or less constant rate. In the leaves, so-called stoma evaporate the sap, and their periodic opening and closing correspond to fluctuating sinks~\cite{Farquhar1974,Katifori2010}. 
Inspired by these natural networks, we wish to address two questions: 
How strong do load fluctuations need to be so that cyclic shunts (loops)~\footnote{In parts of the literature, the term {\em loop} is used synonymous with cycle. Loop, however, may also refer to an edge connecting a node with itself \cite{DiestelBook}.} emerge in the network that break the topology of a spanning tree? 
How far do these cyclic shunts reach into the tree towards the feeding vessel, so that a non-uniform network structure emerges, divided into two subgraphs, one tree-like and the other with cycles? 

The article is organized as follows. In the next section, we introduce the model and simulations that we study. Section 3 and 4 discuss the dynamics on simple network motifs (one source and one sink, and one source and two sinks, respectively). In Section 5 we investigate the emergence of cyclic structures in a larger network with one constant source and many fluctuating loads, which we conclude in the Discussion in Section 6.

\section{Model}

\paragraph*{Network structure.} 
$ V $ denotes the set of nodes of the network with $ N = |V|<\infty $ and $ A \subseteq N \times N $ the set of edges. The edges are bidirectional, so $(i,j) \in A$ implies $(j,i) \in A$. Each node is assigned a pressure $ p_i $. 
The edge flow is $ f_{ij}>0 $ from node $ i $ to $ j $.
Furthermore we assume that the network is resistive and linear, i.e., it is Ohmian with $ f_{ij} = C_{ij} (p_i-p_j) $, where an edge carries the property of a conductance between nodes $ i $ and $ j$ with $ C_{ij}=C_{ji} > 0 $ only if $ (i,j) \in A$.

\begin{figure*}[htp!]
 \centering
 \includegraphics[width=0.8\textwidth]{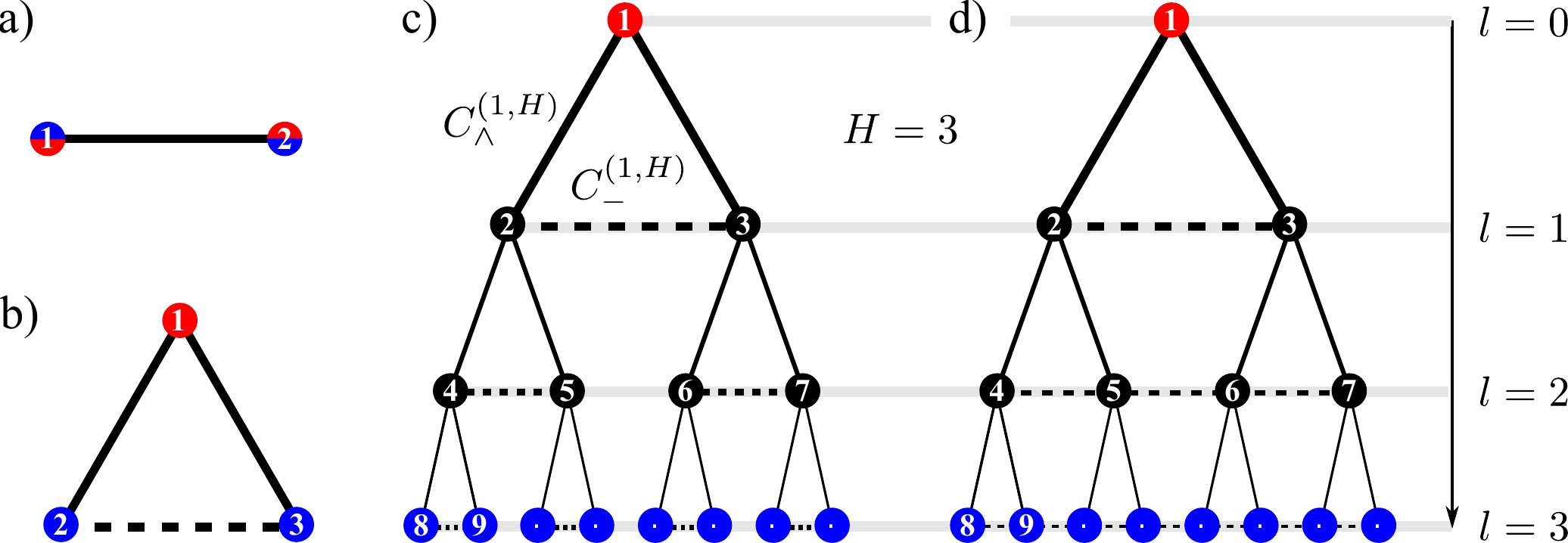}
 \caption{\label{fig:network_topologies} 
 Network structures: (a) motif with 1 fluctuating source and 1 sink (2-node model), (b) motif with 1 constant source and 2 fluctuating sinks (3-node model), and 'augmented trees' (c,d), i.e., tree-like structures with \emph{cross-edges} (dashed horizontal edges) that connect only nodes in each minimal subtree, thus forming a \emph{'triangular tree}' (c), or connect all nodes within one tree level $l$ by a path as in (d). Thus, cross edges introduce cycles to the network. The constant source at the root and the fluctuating sinks in the leaves of the tree are shown in blue and red, respectively, and $l$ denotes the branching level in the tree-like structure.
 }
\end{figure*}

Here, we study the three kinds of wirings illustrated in Fig.~\ref{fig:network_topologies}: a) one source and one sink, b) one source and two fluctuating sinks, c) a tree-like network with height $H$ allowing for cross-edges on every bifurcation/branching level, $l=0,\ldots, H$, leading to cyclic structure. A cycle is a connected subnetwork of $m$ nodes such that each node has exactly two neighbours. 

To model sources and sinks in the network, we include non-zero nodal flows $h_i$. 
Denote by $S \subset V$ the set of sink nodes, $|S| = n = 2^H$. 
We define one source at the edge feeding the network, $h_1 = 1$, and $n$ sinks with $h_i(t)<0$ at all leaves of the tree-like structure (i.e. where capillaries connect to the vein network). For all other nodes, $h_i=0$. 
Mass balance requires that $ \sum_{k \in V} h_k(t) = 0 $ for all $t\in \mathbb{R}$. 

\paragraph*{Mass conservation.}
Mass conservation demands that edge flows, $f_{ij}$ from node $i$ to adjacent nodes $j$, and nodal flows, $h_i$, match the local accumulation rate at node $i$, $dV_i/dt$, i.e.,
\begin{align}\label{massbalance}
 \rho \frac{d}{dt}V_i +  \sum_j f_{ij} &= h_i(t)\
\end{align}
where $\rho$ is the fluid density and $V_i$ is the vessel volume at node $i$.
However, assuming that accumulation is nearly instantaneous or vessels are inelastic the nodal accumulation rate becomes negligible~\cite{Reichholdt2009}.
Mass  balance then becomes Kirchhoff's first law stating that
\begin{align}
 \sum_{j} C_{ij}({p}_i-{p}_j) &=  {h}_i,\
\end{align}
which is re-written in vector/matrix notation by defining the nodal flow  ${\mathbf{ h }} := ({h}_i)_{i \in V}$ and the Kirchhoff matrix ${\mathbf{K}}=({K}_{ij})_{i,j\in V}$ with $K_{ij}:=(\delta_{ij}\sum_j C_{ij})-C_{ij}$,   
\begin{align}\label{eq:Kirchhoff}
 {\mathbf{K}} \cdot {\mathbf{p}} = {\mathbf{h}}\
\end{align}
which is solved for ${\mathbf{ p }} := ({p}_i)_{i \in V}$.

\paragraph*{Dynamically adapting conductances.}
To impose adaptive dynamics to the network, we postulate the generic ad-hoc law for the conductances:
\begin{align}
\frac{d}{dt} C_{ij} &= \alpha_1 C_{ij} (p_j-p_i)^2 - \alpha_2 C_{ij}.\
\end{align}
Thus, the first term on the right hand side induces growth proportional to the power dissipated along the edge, thus mitigating rising pressure differences by increasing the conductance along the edge. Thus, the network adapts itself towards minimizing power consumption. The last term prevents unlimited growth of the conductances. 

Rescaling variables with $ C_{ij}':=h_1^{-1}\sqrt{\alpha_2/\alpha_1}\, C_{ij} $ and $ p_i':=\sqrt{\alpha_1/\alpha_2} \, p_i $, $h':=h/h_1$ (so that $h_1'=1$), $t':=\alpha_2 t$,
the resulting dimensionless model reads 
\begin{align}\label{eq:goveqns}
 \frac{d}{dt'} C_{ij}'(t) &= C_{ij}'(t)[ (p_j'(t)-p_i'(t))^2 - 1],\\\label{eq:Kirchhoff2}
 \mathbf{K}'(t) \cdot \mathbf{p}'(t) &= \mathbf{h'}(t),\\ 
 0&=\sum_{j} h_j'(t).\
\end{align}
where we drop the primes and omit the argument $t$ from now on.

\subsection{Fluctuating sinks}

We consider sinks with periodic and stochastic drive, compliant with $ \sum_{k \in V} h_k(t) = 0 $. Sinks are assumed to fluctuate with a characteristic time scale $T\sim 1/\omega$. Periodic driving may be implemented for $N=2$ with $h_{1,2}=\pm a \cos{\omega t}$ and for $ N=3 $ with $ h_1 = 1 $, $ h_{2,3}=-\half \pm a \cos{\omega t} $. 
For networks with $N>3$ (Fig.~\ref{fig:network_topologies} c)), we implement only stochastic driving.~\footnote {Note that periodic driving may be generalized to larger spanning trees with $ N = 2^L >3$ where $ L $ denotes the number of branching levels: $h_{l} = -\half \pm a^{l {\rm\,mod }\,2}\cos{\omega t}$ where $ l = 2^{L-1}+1, \ldots 2^L $ are indices for the leafs of the tree. Thus, there are $(2^L-2^{L-1})/2$ pairs of leaves that balance each other. To break the symmetry, we may permute the leaf indices $l$ in $h_l$. }

For stochastic driving, let $(s_k)$ be a sequence where for each $k \in \mathbb{N}_0$, the random variable $s_k$ has support $S$ and is distributed identically, uniformly and independently.
Then for time $t$ and sink $i \in S$,
\begin{equation} \label{eq:stochfluc}
  h_i(t) = \begin{cases}
  - \dfrac{1}{n} - \dfrac{a}{\sqrt{2}} & \text{if } i=s_k \text{ with } k=\lfloor t/T \rfloor \\\\
  - \dfrac{1}{n} + \dfrac{1}{n-1}\dfrac{a}{\sqrt{2}}  & \text{otherwise}
\end{cases}
\end{equation}
In plain words, this reflects the situation where at each point $t$ in time, one of the sinks has higher load than the others; after each time interval of length $T$, the sink with higher load is again chosen
uniformly at random. Independent of time, the source (root node) has $h_1 =1$. For all other nodes $j$ (neither source nor sink), $h_j=0$. 

In the system with $H=1$ (one source, two sinks), Eq.~(\ref{eq:stochfluc}) becomes
\begin{equation}
  h_2(t) = \begin{cases}
  - \dhalf - \dfrac{a}{\sqrt{2}} & \text{if } s_k=2 \text{ with } k=\lfloor t/T \rfloor \\\\
  - \dhalf + \dfrac{a}{\sqrt{2}} & \text{otherwise}
\end{cases}
\end{equation}
and $h_3(t) = - 1 - h_2(t)$
a

\subsection{Solving the flow.}
Eqs.~\eqref{eq:goveqns} and \eqref{eq:Kirchhoff2} are invariant with regards to \emph{time-dependent} pressure shifts, $p_k(t)\mapsto p_k(t) + P(t)=:p'_k(t)$. Thus, we may let $P(t):=-N^{-1}\sum_{k\in V}p_k(t)$ and
\begin{align}\label{eq:pinvariance}
 \sum_{k\in V} p'_k(t) = 0,
 \quad \forall\, t\geq 0.\ 
\end{align}
which we later use to obtain~\eqref{eq:Bwedge}.

A general solution of \eqref{eq:Kirchhoff} is of the form $\mathbf{p} = \mathbf{p}_{\rm hm} + \mathbf{p}_{\rm in} \in \mathbb{R}^N$, where $\mathbf{p}_{\rm hm}$ and $\mathbf{p}_{\rm in}$ solve the homogeneous and inhomogeneous problems, respectively.
- Each row in the Kirchhoff matrix $\mathbf{K}\in\mathbb{R}^{N\times N} $ in \eqref{eq:Kirchhoff} has sum zero which implies that $ ( 1,\ldots,1) \in \ker{(\mathbf{K})} $. Since $ {\rm rank}(\mathbf{K}) = N-1 $ (no isolated nodes), we have that $ {\rm ker(\mathbf{K})} = {\rm span((1,\ldots,1))}$ and $ \mathbf{p}_{\rm h} = c\cdot(1,\ldots,1), c \in \mathbb{R} $. In particular, by letting $ c := P(t) $  at any given time $ t\geq 0 $ we may choose a specific instance of the homogeneous solution during the simulation.
- The inhomogeneous solution $\mathbf{p}_{\rm in}$ is determined exactly by solving the reduced system $\mathbf{K}_r \cdot \mathbf{\tilde{p}}_{\rm in} = \mathbf{h}_r$, where the reduced Kirchhoff matrix $\mathbf{K}_r \in \mathbb{R}^{N-1\times N-1}$ with full rank is given by deleting row $l$ and column $l$ in $\mathbf{K}$, and $\mathbf{h}_r \in \mathbb{R}^N{-1}$ is constructed by removing entry $l$ in $\mathbf{h}$. 
Finally, $ \mathbf{p}_{\rm in} $ is given by $p_{{\rm in},l}=0$ and complementing all other entries from $\mathbf{\tilde{p}}_{\rm in} \in \mathbb{R}^{N-1}$.  

\subsection{Simulations}\label{sec:sim}
We use a simple Euler scheme with time step $\Delta t = 10^{-3}$ for numerically integrating Eq.~(\ref{eq:goveqns}).
With a given parameter value $a$, we run the dynamics for a duration of $\tau_\text{sim}=10^4$. 
We take averages and standard deviations of conductances over the time interval [$\tau_\text{sim}/2,\tau_\text{sim}]$.
For a parameter scan, an outer loop runs over values of driving amplitude $a$, starting at $a=1.0$ and decrementing with $\Delta a=10^{-3}$.
For $a=1.0$, the conductance of each edge $(i,j) \in A$ is initialized as $C_{ij}=1$. For $a<1.0$, the integration is initialized with the
conductance averages obtained in the previous run at parameter value $a+\Delta a$.

We have checked that the results are robust under variation of $\Delta t$ and $\tau_\text{sim}$.
In the limit of fluctuations much faster than adaptation, only the distribution values $h_i$ but not their temporal order determines the conductance values obtained. This fact and the symmetry of the tree topology under swapping sink nodes are used to speed up the simulations.

\section{Analysis}
\subsection{One source and one sink (2-node model)}

Let us consider a system with only two nodes $ V = \{1,2 \} $ linked by the edge with conductance $ C_{12} $, as depicted in Fig.~\ref{fig:network_topologies} a).
We assume there are fluctuations driving the system but no net pumping between the two nodes, hence $\langle h_1 \rangle_t = \langle h_2 \rangle_t =0$. 
Invoking Kirchhoff's first law, $C_{12} (p_1-p_2)= h_1(t)$, the model reduces to a single equation,
\begin{align}\label{eq:2node}
 \frac{d} {dt} C_{12} &= C_{12}  \left (\frac{h_1^2}{C_{12}^2}-1\right).\
\end{align}
Superficially, two equilibria seem feasible: (i) the trivial solution with $C_{12} = 0$; and (ii) a non-trivial solution, defined via the condition $C_{12}^2 = h_1^2 > 0$.

To obtain more insight into these solutions, let us assume that the drive $h_1(t)$ has a well-defined characteristic time scale, $T = T(h_1)$. Then we may consider two limiting cases: slow driving ($ T \gg 1 $) and rapid driving ($ T \ll 1 $). 
For slow driving, the conductance is slaved to the driving, i.e., $C_{12}\rightarrow h_1(t)$ as $t\rightarrow \infty$. 
For rapid driving, we may average the equations and seek solutions, $ \langle C_{kl} \rangle $, averaged over rapid fluctuations with characteristic time scale $ T $,
and observe that $ C_{12}^2 \rightarrow \langle C_{12}^2\rangle $ and $ C_{12}^2 \rightarrow \langle h_1(t)^2 \rangle $ as $T\rightarrow 0$.
For slow driving $h_1(t)$ is quasi-stationary and for fast driving $\langle h_1^2 \rangle $ is constant. Therefore, determining stability of the two equilibria is straightforward, as we then simply may inspect the derivative of the right hand side of \eqref{eq:2node}, $-(h_1^2/C_{12}^2 + 1)$. Since $h_1^2>0$ and $C_{12}^2>0$, the non-trivial branch is always stable; however, $C_{12}=0$ corresponds to a singular (and unstable) solution.

For the case of periodic driving of the form
\begin{align}
 h_{1,2}(t) = \pm a \cos{\omega t},\
\end{align}
where $ a\geq 0 $ and $\omega = 2\pi/T$, we may find an explicit (positive valued) solution for Eq.~\eqref{eq:2node}, 
\begin{align}
 C_{12}(t) &= \frac{a}{\sqrt{2}}  
 \sqrt{
     1+\mathcal{T} 
     + \frac{\cos{2 \omega t} + \omega\sin{2\omega t}
    }{1+\omega^2}
 }
\end{align}
with the transient term
$\mathcal{T}:= 2 M e^{-2t}  a^{-2} (\omega^2+1) $ where $M$ is a constant determined by the initial condition.
We may study two different limiting behaviors in the asymptotic limit $t \rightarrow \infty$.
For sufficiently fast driving ($\omega \gg 1$), 
\begin{align}
 C_{12}(t) &\approx \frac{a}{\sqrt{2}}
 \sqrt{1+\omega^{-1}\sin{2\omega t}} 
\
\end{align}
As $ \omega^{-1} \rightarrow 0$, fluctuations in $C_{12}$ become entirely negligible, so that $ C_{12} \rightarrow a/\sqrt{2} $ and $ \langle C_{12}\rangle \rightarrow a/\sqrt{2}$ (see also~\footnote{To be precise, for moderately small $T$, the conductance $C_{12}$  oscillates around $\langle C_{12} \rangle $, with an amplitude that vanishes as $T\rightarrow 0$.}).
For slow driving ($\omega \ll 1$), the conductances are slaved to the driving and we have 
\begin{align}
 C_{12} (t) &= \frac{a}{\sqrt{2}} \sqrt{1 + \cos{2\omega t} }+\bigOh{\omega}.\
\end{align}

\subsection{One source and two sinks (3-node model, $H = 1$)}
We consider a motif with one source with $h_1=1$ and two fluctuating sinks $h_2$ and $h_3$, as depicted in Fig.~\ref{fig:network_topologies} b). For $T\rightarrow 0$ (rapid driving) the sources obey $\langle h_2 \rangle_t - \langle h_3 \rangle_t \rightarrow 0$, i.e., there is no net pumping between nodes $k=2$ and  $k=3$.
The conductances follow the dynamics given by
\begin{align}\label{eq:ds3-1}
 \frac{d}{dt} C_{12} & = C_{12}[(p_1-p_2)^2-1],\\\label{eq:ds3-2}
 \frac{d}{dt} C_{13} & = C_{13}[(p_1-p_3)^2-1],\\\label{eq:ds3-3}
 \frac{d}{dt} C_{23} & = C_{23}[(p_2-p_3)^2-1],\
\end{align}
with
\begin{align}\label{eq:mb3-1}
 1 & = C_{12} (p_1-p_2) + C_{13} (p_1-p_3),\\\label{eq:mb3-2}
 h_2 & = C_{12} (p_2-p_1) + C_{23} (p_2-p_3),\\\label{eq:mb3-3}
 h_3 & = C_{13} (p_3-p_1) + C_{23} (p_3-p_2).\
\end{align}
Eight (quasi-stationary) solutions are conceivable where either $C_{kl}>0$ or $C_{kl}=0$ for every edge $(k,l)$; however, only solutions with 
$C_{12}>0$ and $C_{13}>0$ are physically meaningful, and so, only the two solutions with $C_{23} = 0$ or $C_{23}>0$ are feasible. 
Note that unlike the case of the 2-node motif, the branch $C_{23}=0$ is not singular anymore. 

The general solution for $N \geq 3$ nodes is more intricate than for $N=2$ nodes. Therefore, we limit our analysis from now on to the case of rapid driving, where $T \ll 1$ is very small, and we consider only  the asymptotic solutions where $ t\rightarrow \infty $. 
In analogy to the 2-node motif, we seek solutions, $ \langle C_{kl}\rangle $, averaged over rapid fluctuations with characteristic time scale $T$.
Considering  that $C_{ij}$ changes on a slow time scale, $ C_{ij} \rightarrow \langle C_{ij} \rangle$ as $T \rightarrow 0 $, and therefore we may from now on use $\langle C_{kl}  \rangle$ and $ C_{kl} $ interchangeably and omit $\langle \cdot \rangle $ around conductances.
For symmetry reasons, rapid driving implies that $\langle C_{12}\rangle = \langle C_{13}\rangle$.
In this limit, the dynamics of the conductances is then constrained to a two dimensional symmetry manifold and effectively reduced to the two equations given by
\begin{align}\label{eq:ds32-1}
 \frac{d}{dt} C_{12} & = C_{12}[\langle(p_1-p_2)^2\rangle-1],\\\label{eq:ds32-2}
 \frac{d}{dt} C_{23} & = C_{23}[\langle(p_2-p_3)^2\rangle-1],\
\end{align}
together with Eqs.~\eqref{eq:mb3-1}-\eqref{eq:mb3-3}, in which $h_k$ and $p_k$ fluctuate rapidly but $C_{kl}$ may be considered quasi-stationary. Pressures may be eliminated by observing the following equalities. Subtracting \eqref{eq:mb3-2} from \eqref{eq:mb3-3}, we have
\begin{align}\label{eq:p2p3}
 p_2-p_3 &= \frac{h_2-h_3}{C_{12}+2C_{23}}, \
\end{align}
and substitution of this expression into \eqref{eq:mb3-2} yields
\begin{align}\label{eq:p1p2}
 p_2 - p_1 &=  \frac{1}{C_{12}}\left [h_2 - (h_2-h_3)\frac{C_{23}}{C_{12}+2C_{23}}\right ].\
\end{align}

We first consider the 'tree-like' solution branch with a non-conducting cross edge $C_{23}=0$. Stationarity for $C_{12}>0$ implies that $\langle (p_1-p_2)^2\rangle=1$, and mass balance \eqref{eq:p1p2} requires that $p_2-p_1 = h_2/C_{12}$. Therefore, the tree-like branch is given by 
\begin{align}
 \Bwedge &= (C_{12},C_{23}) = \left ( \sqrt{ \langle h_2 ^2 \rangle},0 \right ).\ 
\end{align}

\begin{figure}[htp!]
 \centering
 \includegraphics[width=0.45\textwidth]{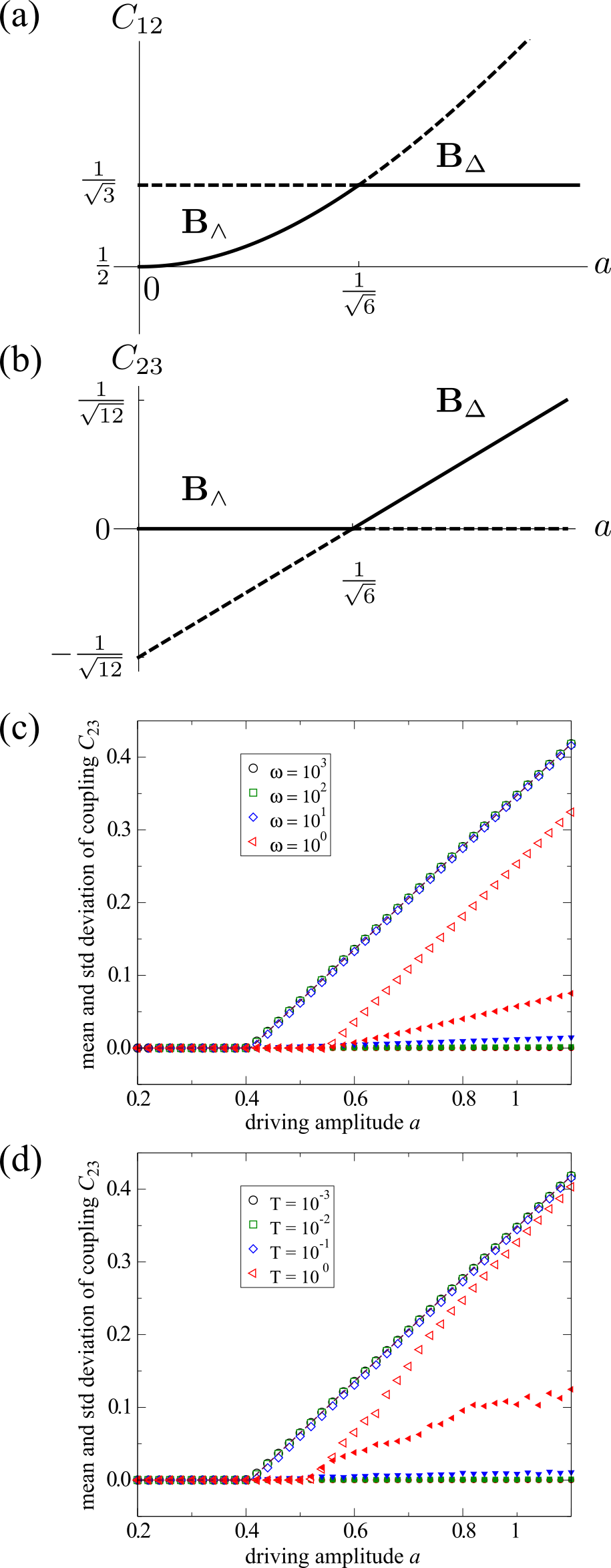}\\
 \caption{\label{fig:3nodemodel} 
 Bifurcation diagrams for the 3-node model (Fig.~\ref{fig:network_topologies} (b)). Analytical solutions for periodic driving are shown in (a) and (b).
 Simulated behavior for the shunting conductance $C_{23}$ is shown for periodic driving (c) and rapid stochastic driving (d).    
 Mean values (large open symbols) and standard deviation (small filled symbols) are measured for $C_{23}(t)$ over the entire simulation.  
 For both periodic and stochastic driving the predicted bifurcation point $a_{\rm c} \rightarrow 1/\sqrt{6}$ is asymptotically reached  as the characteristic driving frequency $T^{-1}\rightarrow \infty$. 
}
\end{figure}

Second, we consider the non-trivial branch with $C_{23}>0$.
Again, we first look for a stationary solution $\langle C_{12} \rangle>0$ implying that $1 = \langle (p_1 - p_2)^2 \rangle$.
Using the symmetry $\langle C_{12}\rangle = \langle C_{13}\rangle$ and \eqref{eq:mb3-1} allows us to write $ 1 = C_{12} (- p_2 - p_3 + 2 p_1) $. 
Eq.~\eqref{eq:pinvariance} implies $-p_2-p_3 =  p_1$, and we obtain the constant pressure $ p_1^* =  1/(3C_{12})$ in node $k=1$.
Next, Eq.~\eqref{eq:pinvariance} implies also $p_1-p_2 = 3/2 p_1 -1/2(p_2-p_3)$. Thus, $1=\langle (p_1-p_2)^2\rangle = 9/4 \langle p_1^2 \rangle - 3/4 \langle p_1(p_2-p_3) \rangle -1/4 \langle (p_2-p_3)^2 \rangle = 9/4 (p_1^{*})^2 + 1/4$, since by assumption $(C_{12} + 2 C_{23}) \langle p_2 - p_3 \rangle = \langle h_2 - h_3 \rangle = 0$. Hence, we have $ p_1^* =1/\sqrt{3} $ and therefore $C_{12} = 1/\sqrt{3}$. 
Finally, to determine $C_{23}$ we use \eqref{eq:p2p3} and $\langle (p_2-p_3)^2 \rangle = 1$. We obtain
\begin{align}\label{eq:Bwedge}
 \Btriangle &=(C_{12},C_{23}) = \left (\frac{1}{\sqrt{3}}, \half \left [\sqrt{\langle (h_2-h_3)^2 \rangle } - \frac{1}{\sqrt{3}} \right ] \right). \
\end{align}
For periodic driving we let $ h_{2,3} = -\half \pm a \cos{\omega t} $ and obtain the solutions
\begin{align}\label{eq:3node_wedge}
 \Bwedge^{\rm per} &= \left( \sqrt{\frac{1}{4}+\half a^2},0 \right),\\\label{eq:3node_triangle}
 \Btriangle^{\rm per} &= \left(\frac{1}{\sqrt{3}} ,\frac{1}{\sqrt{2}} \left ( a - \frac{1}{\sqrt{6}} \right) \right).\
\end{align}
From $\mathbf{B}_\triangle^{per}$ we read off the critical value for the drive amplitude, $a_{\rm c}= 1/\sqrt{6}$, at which $C_{23}>0$ becomes physically viable.

We now study stability for $\Bwedge$ and $\Btriangle$ by considering Eqs.~\eqref{eq:ds32-1} and \eqref{eq:ds32-2} which restrict dynamics to the two dimensional subspace defined by $C_{12}=C_{23}$. Using \eqref{eq:p1p2} and \eqref{eq:p2p3} and eliminating pressures, we may compute the Jacobian which we evaluate for the two branches to obtain the corresponding eigenvalues for periodic driving,
\begin{align}
  \lambda_1^\wedge & = \lambda_1^\triangle = -2,\\
  \lambda_2^\wedge & = \frac{6 a^2-1}{2 a^2+1},\\
  \lambda_2^\triangle & = \frac{3 }{2 }\left(\frac{1}{\sqrt{6}}-a\right).\
\end{align}
More generally, we find $\lambda_2^\wedge = (\langle h_3^2\rangle -2\langle h_2h_3\rangle )/ \langle h_2^2\rangle$, but $\lambda_2^\triangle$ yields an unwieldy expression.
Thus, the two branches swap stability  in a transcritical bifurcation at $a_{\rm c}=1/\sqrt{6}$, so that $\Bwedge$ is stable for $a<a_{\rm c}$ and $\Btriangle$ is stable for $a>a_{\rm c}$, as shown in Fig.~\ref{fig:3nodemodel}, a) and b). In panels c) and d) of Fig.~\ref{fig:3nodemodel} we show simulations for periodic and stochastic driving with varying time scales $T$. These results demonstrate that the simulated behavior converges to our analytical predictions as the characteristic driving frequency $T^{-1}\rightarrow \infty$.


\subsection{One source and multiple fluctuating sinks (augmented tree, $H > 1$)}
\label{sec:tree}
\begin{figure*}
 \centering
 \includegraphics[width=0.9\textwidth]{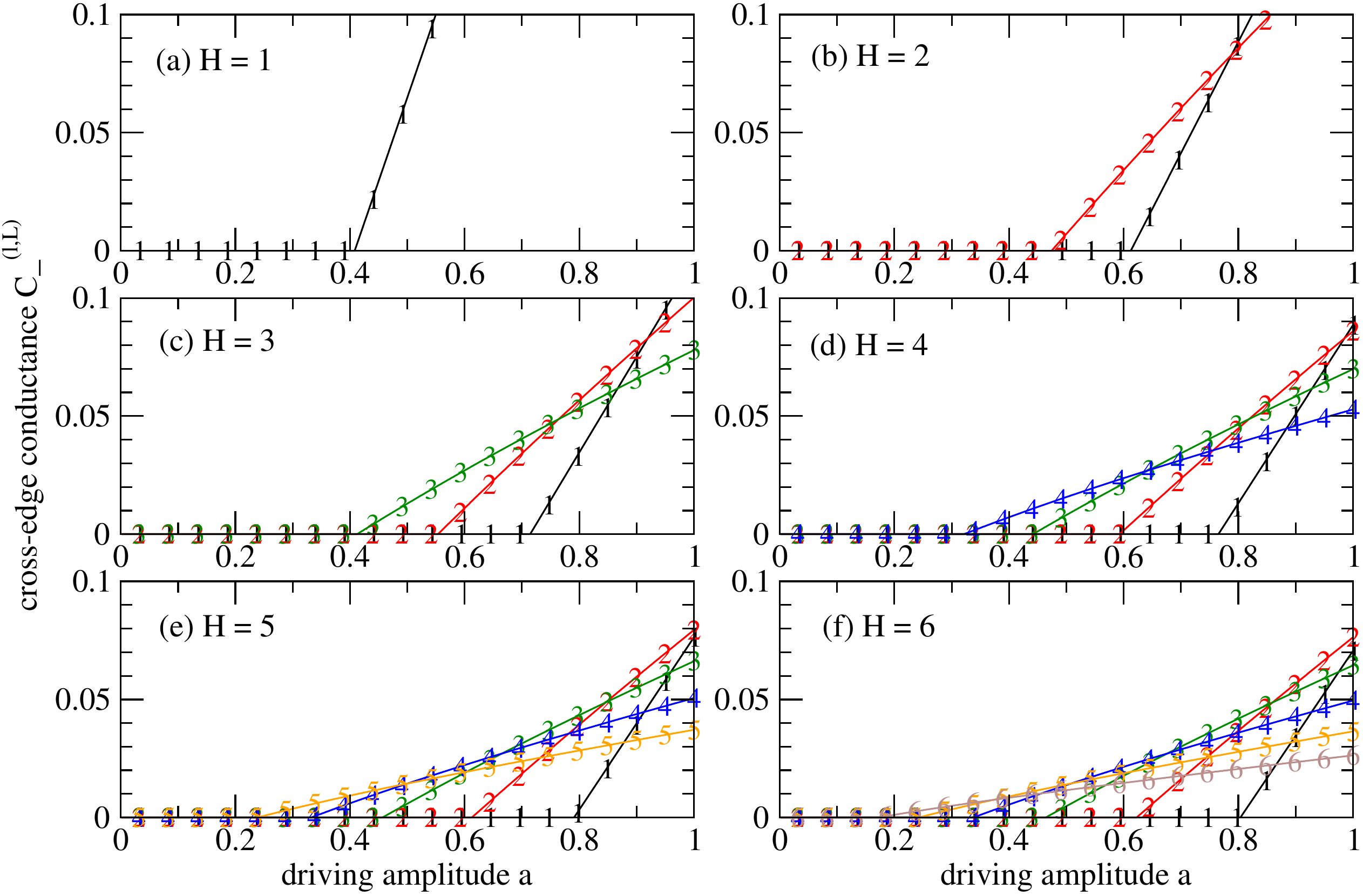}
  \caption{\label{figuren} 
  Bifurcation diagrams for augmented tree systems ($H>1$) according to Fig.~\ref{fig:network_topologies} c) with dynamics under fast stochastic driving at the sinks. The conductances of cross-edges are shown for different levels $l$ as defined in Fig.~\ref{fig:network_topologies}, i.e., from the root ($l=0$) to the sinks in the leaves ($l=H$). The values result from parameter scans decreasing driving amplitude $a$ from 1 to 0. Panels (a)-(f) distinguish systems with different heights $H \in \{1,2,\dots,6\}$. In each case, the number of nodes is $N=2^{H+1}-1$, i.e., $N=127$ nodes for $H=6$.}
\end{figure*}

\begin{figure*}
 \centering
 \includegraphics[width=0.8\textwidth]{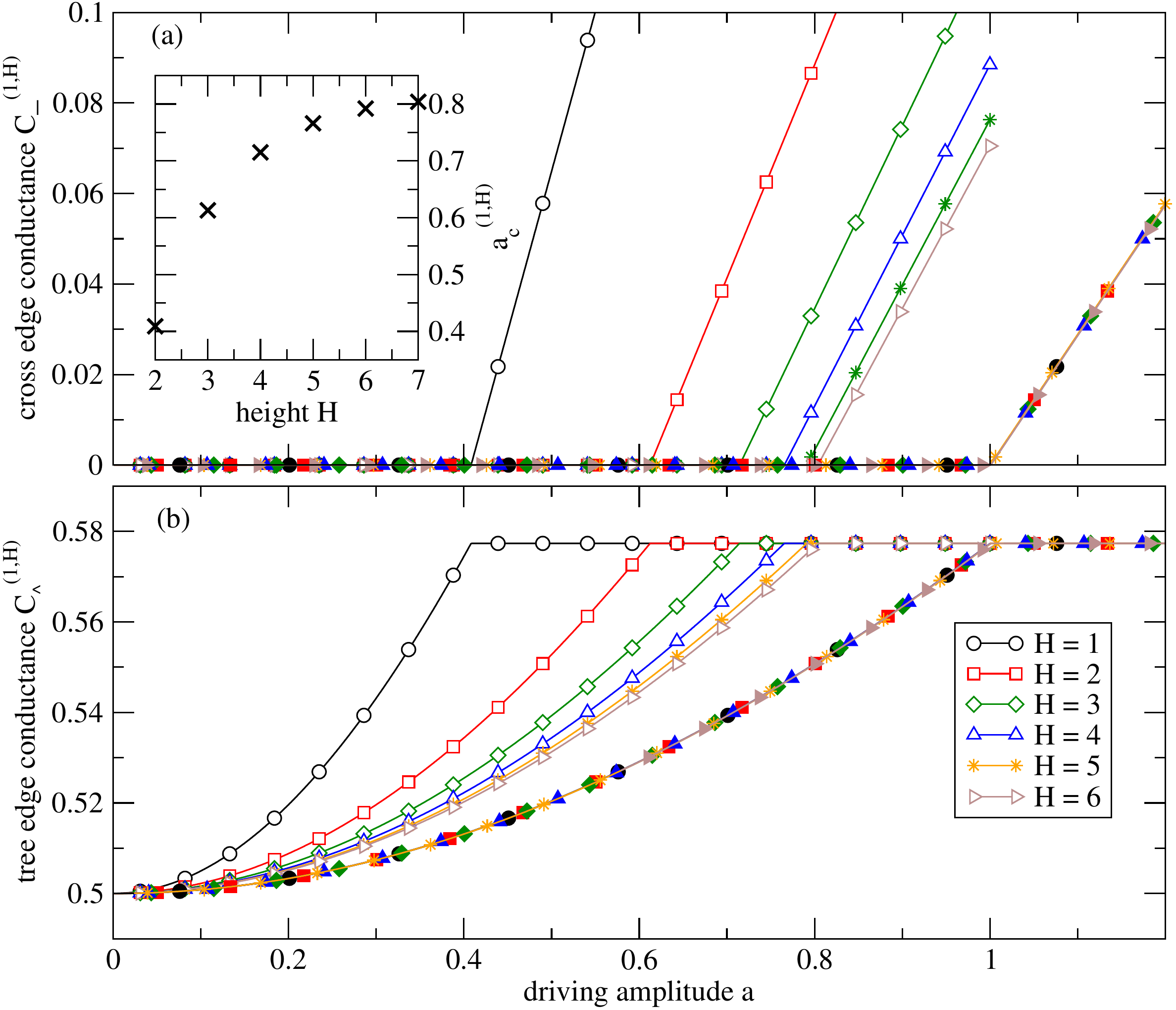}
  \caption{\label{figurenrescaled} 
  Parameter dependence of stationary conductances in the top triangle of systems with height $H$.
  Open symbols denote in panel a) the conductance of the root's tree edges, $C_{\wedge}^{(1,H)}$, and in panel b) the conductances of the root's cross-edge, $C_{-}^{(1,H)}$. Filled symbols are the same values plotted as a function of $a$ rescaled with the critical value $a_{\rm c}^{(1,H)}$. 
  The inset in panel a) shows the critical values $a_{\rm c}^{(1,H)}$. For height $H$, the system has $N=2^{H+1}-1$ nodes, i.e., $N=127$ nodes for $H=6$.
  }
\end{figure*}

\begin{figure*}
  \centering
  \includegraphics[width=0.8\textwidth]{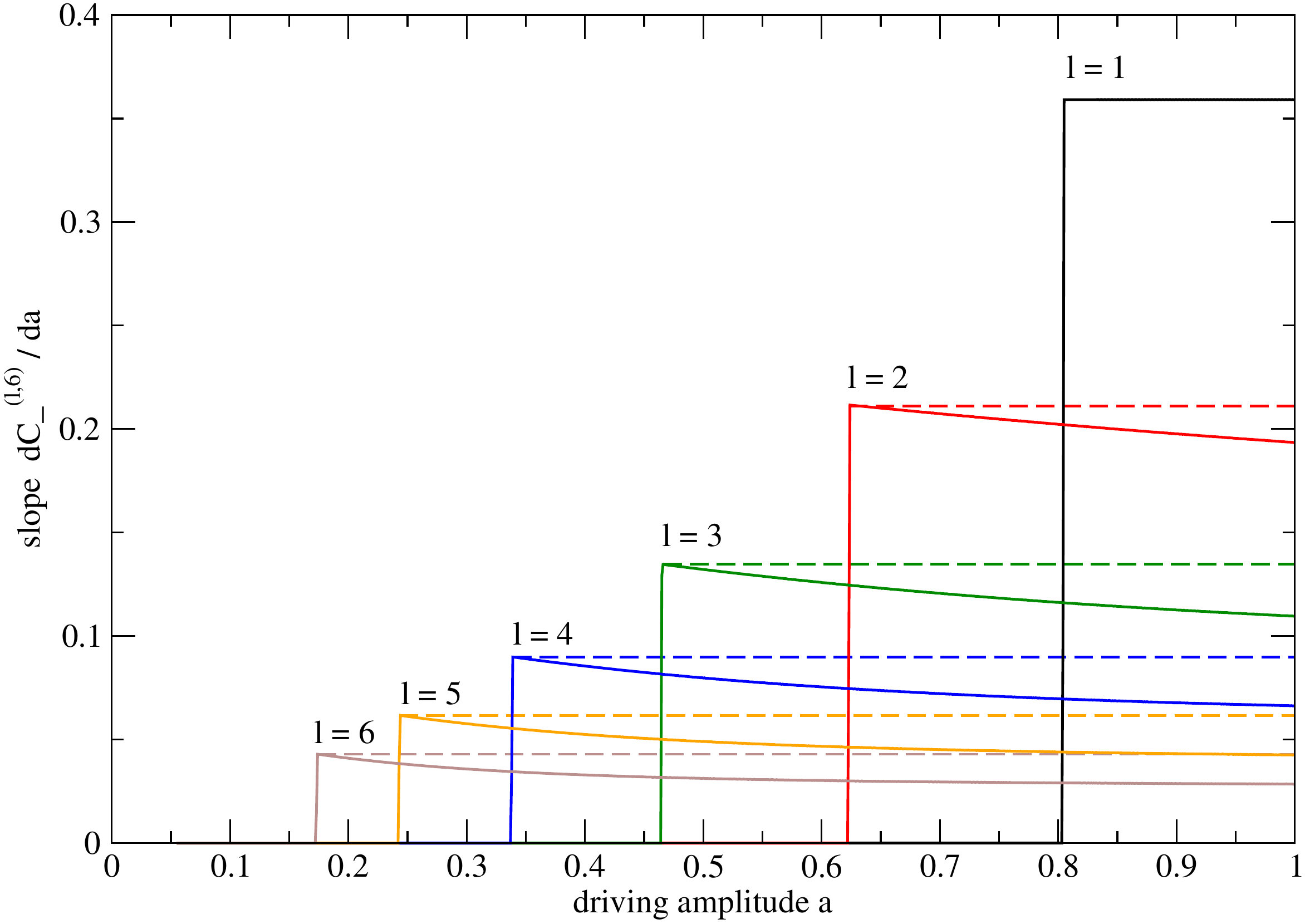}
  \caption{
  \label{figurend} 
  Slopes of conductances of cross edges in the system with height $H = 6$ with $n=64$ sinks and $N=127$ nodes in total.
  The plotted curves are the numerically determined derivatives of the data plotted in Fig.~\ref{figuren} f). For each curve, the maximum value is indicated by an extra vertical dashed line.
  Note that the derivative is piecewise constant only for the cross edge closest to the root. The other conductances exhibit maximal slope only at the critical point.
  }
\end{figure*}

Let us now turn to larger systems with more than two sinks and nodes representing intermediate branching points between source and sinks.
The networks considered are complete binary trees, augmented with cross edges directly at each branching, shown as dashed edges in Fig.~\ref{fig:network_topologies} c). We consider the stationary conductance values under stochastic driving with a rapid characteristic time scale $T\ll 1$.

For a level $l\in\{0,1,\dots,H\}$ in a system with height $H$, we consider the dynamics for conductances of \emph{tree-edges}, $C_\wedge^{(l,H)}$, and of \emph{cross-edges}, $C_-^{(l,H)}$, see Fig.~\ref{fig:network_topologies} c). 
By $a_{\rm c}^{(l,H)}$ we denote the critical driving amplitude above which cross-edges with non-zero conductances emerge.
Fig.~\ref{figuren} shows the asymptotic values of the conductances of cross-edges for fluctuations with varying amplitude $a$. Non-zero conductances in cross-edges appear with increasing amplitude in an order that follows the hierarchy of the tree, i.e., $a_{\rm c}^{(k,H)} > a_{\rm c}^{(l,H)}$ for all $0\le k < l < H$.  To illustrate, as we increase $a$, the cross-edges between sinks, $l=H$, are the first that begin to conduct. Then as we increase $a$ even further, the next level, $l = H-1$, obtains conducting cross-edges, and so on.
Accordingly, the last transition to non-zero conductance for increasing amplitude $a$ occurs in $C_{-}^{(1,H)}$ between the root's child nodes. 

One may speculate if all observed transitions from zero to non-zero conductance are of the same kind at different levels, only varying in the parameter value $a_{\rm c}^{(l,H)}$ at the transition and the slope in the supercritical regime. Fig.~\ref{figurenrescaled} shows that this is indeed the case for the cross-edge at level $l=1$. 
Plotting the conductances of the edges of the source node as a function $a/a_{\rm c}^{(l,H)}$, we observe a perfect collapse of all data for varying $H$ (and $ l=1 $  fixed), by rescaling $ a \mapsto a / a_{\rm c}^{(l,H)} =: \tilde{a} $ in systems of different heights $ H $. So the
top triangle including the source node of an augmented tree of general height $ H $ behaves in the same way as the $ H = 1 $ system with driving parameter $ a $ rescaled. Thus, the analytic results \eqref{eq:3node_wedge} and \eqref{eq:3node_triangle} for $ H = 1 $ allow us to express the scaled conductances explicitly,
\begin{align}
 \tilde{C}_{12}\big|_{l=1} &= \dhalf \sqrt{1+\frac{1}{3}\tilde{a}^2},\\
 \tilde{C}_{23}\big|_{l=1} &= \frac{1}{\sqrt{12}} (\tilde{a}-1).\
\end{align}

The $a$-dependence of conductances on the other levels is more intricate, as is shown in Fig.~\ref{figurend} for the case of a system with $H = 6$ levels.
The derivatives of cross-edge conductances reveal detail not apparent in the coarser plot (Fig.~\ref{figuren}) of these conductances themselves.
The slope is maximal at the transition (as the one-sided derivative with $a$ approaching $a_{\rm c}(l,6)$ from above). 
As $a$ is increasing, the conductance curve becomes slightly flatter. 
This non-linear effect hints at dependencies between the levels, $l$.
However, one may show that using the same rescaled amplitude parameter, $\tilde{a}$, $C_{\wedge}^{(l,H)}$ and $C_{-}^{(l,H)}$ collapse to single curves for fixed $l$ while varying $H$ (not shown).

The Supplementary Material of this article provides four videos of simulations of the dynamics for systems with height $H=4$. Each of these shows a parameter sweep as described in Section~\ref{sec:sim}.
Video 1 ({\tt Anc\_Video\_1.gif}) shows the dynamics for the network structure studied above, i.e., with one cross edge at each branching as illustrated in Fig.~\ref{fig:network_topologies} c). 
Video 2 ({\tt Anc\_Video\_2.gif}), the tree is further augmented to allow for cross edges bridging longer distances in the tree; such a network structure is shown in Fig.~\ref{fig:network_topologies} d).
Two further Videos 3 and 4 ({Anc\_Video\_3.gif}, {\tt Anc\_Video\_4.gif}) employ damaged versions of the latter structure in which roughly half of all cross edges have been randomly selected and removed. 
These simulations confirm the observation that the cross edges begin to conduct in order of the tree hierarchy, starting from sinks and moving towards the source.


\section{Discussion and Outlook}

We have studied how fluctuating loads affect the re-configuration of vessels in an adaptive flow network. To do this, we have introduced a minimal model, consisting of a resistive network with conductances on the edges which adapt dynamically towards minimizing fluctuations in the network. To be explicit, the conductance is up-regulated with the pressure drop squared (power dissipated), but down-regulated with a factor proportional to the conductance. Although adaptation adheres to local rules they also experience a global feedback through the coupling via the flow network. An important question, that also appears in the context of synaptic plasticity~\cite{Tetzlaff2011}, is then what are the (stable) equilibrium configurations of conductance in the network.
Furthermore, we introduced load fluctuations by including fluctuating sinks in certain network nodes. Assuming that these fluctuations own a well defined characteristic time scale much more rapid  ($T\rightarrow \infty$) than the time scale ($\sim 1$) of adapting conductances allowed us to treat fluctuations in terms of their averages over time, i.e., their amplitudes. Considering both periodic and stochastic fluctuations, we analytically and numerically investigated small and large networks to determine how their equilibrium conductance configurations depends on the amplitude of the load fluctuations, $a$. In particular, we used this model to investigate how far into the network the fluctuations would be able to induce re-configurations.

First, we investigated two very simple network motifs regarding the space of equilibrium configurations and their stability. 
The dynamics for the motif consisting of one fluctuating sink and source can be solved exactly (Fig.~\ref{fig:network_topologies} a)). The conductance connecting the two nodes is non-zero and stable for all fluctuation amplitudes with $a>0$. The triangular motif with one constant source and two fluctuating sinks (Fig.~\ref{fig:network_topologies} b)) can also be solved analytically but exhibits a transcritical bifurcation at a critical drive amplitude $a_c$: for sub-critical drive, $a<a_c$, the tree-like solution branch without cross edge is stable; for super-critical drive, $a>a_c$, the cyclic structure (triangle with non-zero tree-edge and cross-edge) is stable, see Fig.~\ref{fig:3nodemodel}. Numerical simulations demonstrate that the system behavior asymptotically approaches the solutions obtained for the limit of rapid driving ($T\rightarrow 0$) as $T\rightarrow \infty$.

Next, we ran numerical simulations of larger tree-like networks with a constant source at its root and fluctuating loads in the leaves. 
Using an initial configuration $C_{ij}|_{t=0}=1$ for all edges $(i,j)\in A$, we then investigated into which quasi-stationary configuration the network settles. To investigate the system behavior analytically, we studied two scenarios.  
-- First, we assumed a simplified network structure, i.e., a complete binary tree with cross-edges only at each branching, see Fig.~\ref{fig:network_topologies} c) and Ancillary (supplementary) Video 1. 
As the amplitude $a$ exceeds a critical threshold $a_{\rm c}(l,H)$, conductances $C_-^{(l,H)}$  of cross-edges between sinks on level $l$ transition from a zero to a non-zero positive value. 
The critical driving amplitudes $a_{\rm c}(l,H)$, depending on the edge level $l$ and height $H$ of the tree-like structure, are hierarchically ordered so that the transitions to non-zero cross-edge conductances appear successively in descending order of $l$ as the amplitude grows.
Thus, we observe that, as the amplitude is increased, cycles first emerge near the fluctuating sinks and then spread towards the root with the constant source. 
-- Second, we compared these observations with the behavior in a generalized network topology. In this second scenario, we ran numerical simulations of networks lacking intrinsic tree structure, i.e., where all cross-edges are allowed, see Fig.~\ref{fig:network_topologies} d) and Ancillary Video 2. In this case, the threshold for an edge $(i,j)$ between sinks $i$ and $j$ has a dependency on the length of the tree path between $i$ and $j$. The longer the tree path (distance on tree), the lower its threshold. This means that -- as the amplitude is increased -- 
minimal distance short-cuts for longer paths form before the cross-edges from the first scenario.

In both scenarios, however, the network is partitioned into two regions with tree-like motifs near the source at the root and cycles closer to the fluctuating sinks in the leaves of the tree. Transitions get more complicated when randomly chosen cross-edges are topologically deleted (Ancillary Videos 3 and 4).

Other work has investigated the appearance of cyclic structures (loops) by discussing flow networks via optimization in terms of minimizing dissipative losses under continuous reconfiguration of the flow conditions~\cite{Katifori2010,Dodds2010,Corson2010}. While this point of view may be motivated by evolutionary principles and yielded many interesting insights regarding the critical emergence of cycles, we were interested in formulating a minimal dynamical model~\cite{Grawer2015}. Contrasting previous studies, we did not assume spatially uniform fluctuations and investigate conditions under which cyclic structures form in general; rather, we investigated if partitions between tree-like and cyclic structures emerge would form and where their boundaries lie. Many studies in vascular physiology focused on the dynamics of single vessels and have complex biophysical models including a large degree of complexity prohibitive to mathematical analysis, and only few have computationally investigated dynamics in networks~\cite{Hacking1996,Postnov2016}; here, we tried to systematically address stability of such vessel models from a mathematical perspective.

The study on simple network motifs has been insightful and complemented the numerical findings that we have made for larger trees. Further research may address equilibrium configurations and their transitions ($a_c^{(l,H)}$) from tree-like to cyclic structures in network structures shown in Fig.~\eqref{fig:network_topologies} c) and d), and attempt to find complete solutions via mathematical analysis and scaling arguments. For the case of more general network topologies (see e.g., Fig.~\eqref{fig:network_topologies} d) and for the case of heterogeneities, an open question remains whether multi-stable equilibria are possible. Furthermore, research should be conducted on reducing complex biophysical dynamics to simpler mathematical models, which could be investigated towards identifying classes of different dynamic behavior. Finally, generalizations to co-evolving networks with edges being dynamically created/deleted may be considered~\cite{TaylorKing2016}.

%


\section*{Funding}
We acknowledge travel funding from  Action CA15109, European Cooperation for Statistics of Network Data Science (COSTNET).
Research conducted by EAM is supported by the Dynamical Systems Interdisciplinary Network, University of Copenhagen.
KK acknowledges funding from MINECO through the Ram\'{o}n y Cajal program and through project SPASIMM, FIS2016-80067-P (AEI/FEDER, EU).

\section*{Acknowledgments}
EAM would like to thank J.~C. Brings Jacobsen for helpful discussions on circulatory physiology and E.~Katifori on adaptive networks. 

\section*{Ancillary (supplementary) files }
Ancillary files for this article consists of 4 Ancillary Videos (1-4) which are described at the end of Section~\ref{sec:tree}. A legend for the conductances is supplied in Image 1 {\tt Anc\_legend\_for\_films.png}.

\bibliographystyle{apsrev4-1}
\bibliography{references}

\end{document}